\documentclass{aa}
\def\Teff{{$T_{\rm eff}$}~}
\def\Msun{{$M_{\sun}$}~}
\def\Mbol{{$M_{\rm bol}$}~}

\def\dydz{{$\Delta Y/\Delta Z$}~}
\def\Yp{{$Y_{\rm p}$}~}

\usepackage{graphics}

\begin{document}

\thesaurus{08(08.01.1;08.08.1;08.09.3;08.12.1;10.01.1;10.19.1)}



\title{The Hipparcos HR diagram of nearby stars in the metallicity range:
-1.0 $<$ [Fe/H] $<$ 0.3}

\subtitle{A new constraint on  the theory of  stellar interiors
and model atmospheres}

\author{Y. Lebreton\inst{1}\thanks{e-mail address: Yveline.Lebreton@obspm.fr}
\and M.-N. Perrin \inst{2}
\and R. Cayrel \inst{2}
\and A. Baglin \inst{3}
\and J. Fernandes \inst{4}}

\offprints{Y. Lebreton}

\institute{DASGAL, CNRS URA 335, Observatoire de Paris, Place J.  Janssen, 92195
Meudon,  France  \and  DASGAL, CNRS URA  335,  Observatoire  de  Paris, 61 Av.  de
l'Observatoire, 75014 Paris, France \and DESPA, CNRS URA 264, Observatoire de Paris,
Place J.  Janssen, 92195
Meudon,  France  \and Observat\'orio Astron\'omico da
Universidade de Coimbra, 3040 Coimbra, Portugal }

\titlerunning{The  HR  diagram  for  late-type  nearby  stars  after  Hipparcos}
\authorrunning{Y. Lebreton et al.}

\date{Received /Accepted }

\maketitle

\begin{abstract}

The  Hipparcos mission has provided very high quality parallaxes of a
sample  of  a  hundred nearby  disk  stars,  of spectral types  F  to  K.
In parallel, bolometric fluxes, effective temperatures, and accurate Fe/H
ratios of many of these stars became available
through infrared photometry and detailed spectroscopic analyses.
These new accurate data allow to build the Hertzsprung--Russell 
diagram of stars of the solar neighbourhood with the smallest error bars
ever obtained. 

We analyse these observations by means of
theoretical stellar models, computed with the most recent input physics.

We first examine the positions of the objects versus standard theoretical
isochrones, corresponding to their che\-mical composition and age.
For these isochrones we have first assumed that the helium content
was varying in locksteps with metallicity.
The comparison becomes age-inde\-pen\-dent in the lower part of the HR diagram,
where evolutionary effects are negligible. We show that for the unevolved
stars, the agreement between real stars and models is fairly satisfactory
for stars
with metallicity within $\pm $ 0.3 dex of the solar metallicity, but that a conflict
exists for stars with metallicity less than [Fe/H] = $-0.5$. This conflict
cannot be resolved by decreasing the helium abundance: values of this
abundance below the primordial abundance would be required.

On the basis of recent works, we show that the addition of two processes 
not included in standard models  can help solving the above discrepancy. These are
 (i) correcting the LTE iron abundances using
a non-LTE approach and (ii) including microscopic diffusion of He and heavier 
elements in the stellar interior. 
 The case of the binary star $\mu$~Cas is
particularly useful to support this conclusion as its mass is also known
from its orbit. After inclusion of the two effects, $\mu$~Cas~A falls on its 
expected isochrone, within the error bars corresponding to its mass.

All stars with -0.3 $< $ [Fe/H] $<$ 0.3 are located between the helium-scaled 
isochrones corresponding to these metallicities. However five of them
are not located exactly where they are expected to be for their 
metallicity. This may reflect a helium content lower than the metallicity-scaled
value. But not necessarily, as a possible sedimentation of the elements  
might complicate the determination of the helium content .
 The age of main sequence  solar composition stars covers a large range,   
 and the effects of sedimentation are time dependent.

\keywords{stars:  abundances -- stars:  HR diagram -- stars:  interiors --
          stars: late-type --  Galaxy: abundances -- Galaxy: solar neighbourhood}

\end{abstract}

\section{Introduction}

Valuable tests of the stellar evolution theory and strong
constraints on the physical description of stellar interiors
are mainly provided by the study of the most-accurately observed
stellar objects. The number of observable parameters accessible
through observations is important as well as the accuracy on
their determination. The Sun, the nearby stars including 
members of visual binary systems, and stars belonging to
open clusters represent some of these best-known objects.

The observations of nearby low-mass stars have been greatly
improved,  and their number has substantially increased, 
during the last few years. Distances have been
determined with a very high precision with the Hipparcos satellite.
In parallel, ground-based measurements have provided high resolution
spectra  and  multicolour  photometry  of  an appreciable number of stars
of the solar neighbourhood. On the other hand, models
of stellar  atmospheres have largely benefited from progresses in
the theoretical description of microscopic physics, in particular
opacities. The analysis of the observational data, using
model atmospheres, has provided bolometric magnitudes, effective
temperatures and abundances with an accuracy which had never been
reached previously. In addition, several nearby stars are members
of visual binary systems and their mass is known. 

The study of stars of low mass has many important physical and
astrophysical implications. First, several uncertainties remain
on the physics of their internal structure. For instance,
the unperfect understanding of the convective transport or of the
non local thermodynamical equilibrium (non-LTE) effects in
atmospheres and envelopes have important consequences: atmospheres
and envelopes serve as external boundary conditions for the
interior models, fixing the radius, and are also
used in the analysis of observational data to
determine abundances, effective temperatures and gravities.
On the other hand, transport processes at work in the
deep interior need to be better constrained; they link
the convective zone to deep nuclear burning regions, involving
consequences for the surface abundances (lithium and
light elements but also heavy elements), and for the estimated age
and global parameters, when fresh helium is brought to the burning
core.

 An important question is also that of
helium content, not measurable in the photosphere
due to the lack of lines in the spectrum. The initial helium
content of a star determines its lifetime and internal structure and
is also a witness of past galactic history.
The knowledge of the initial helium abundance of stars born in
different sites with different metallicities is
therefore fundamental for studies of the chemical evolution of the Galaxy.  

We study here a sample of a hundred nearby disk stars, of
spectral types in the range F to late K, by means of theoretical
stellar models. Our aim is to discuss the ability of the models
to reproduce high quality observations and, when possible, to
determine the helium content of the stars.

In Sect.~\ref{previous-results} we make a rapid summary of some
results previously obtained on the subject, from studies of
a few particularly well-known objects.
In Sect.~\ref{observations} we carefully
examine  the  observational  data  available  for nearby stars
in  order  to extract subsamples corresponding  to  the
very  best  accuracies. We then present the derived
Hertzsprung--Russell (HR) diagrams of stars of the solar
neighbourhood which have the smallest error bars ever obtained. Section~4
describes the theoretical models used to compute isochrones for given
chemical compositions. In Sect.~5
we show evidence for discrepancies between observations and 
theoretical models, which are particularly obvious for stars with
metallicities below [Fe/H] =-0.5.
 In Sect.~6 we examine this
problem in the light of recent works on departure from LTE for iron 
(Th\'evenin \& Idiart \cite{TI99}) and on microscopic diffusion of helium
and heavier elements in metal-poor stars (Morel \& Baglin \cite{MB99}).
We show that the cumulative effects of these two processes is large
enough to remove the discrepancy shown in Sect.~5. Section~7 considers 
the case of binary stars with known dynamical masses. Section~8 adresses
the question of the helium content of individual objects of the sample.
Section~9 summarizes our conclusions. 

\section{Present status of helium content and mixing-length parameter determinations}
\label{previous-results}

Since observations do not provide direct
determinations of helium, assumptions have to be made for
the initial helium content of models of low-mass
stars. Very often it is supposed that the metallicity Z and
helium Y are related through a constant \dydz-value.

\dydz = ($Y$ - \Yp)/$Z$ is often determined from the results of the solar
calibration and for any other star of known metallicity, the
helium abundance is scaled on the solar one. 
In addition of the behaviour of Y with Z, another problem is the universality
of the $\alpha = l/H_p $ ratio of the mixing length 
over the pressure scale-height in the mixing-length treatment of the 
convection.
We now briefly
review the various determinations of these parameters, in the closest stars
and various astrophysical sites.

\subsection{ Nearby stars}

Perrin et~al.  (\cite{per77}) were the first to examine the HR diagram of a
selected sample of the nearest low-mass stars.  Because in 1977 the error bars
in  the HR diagram were quite large, Perrin et~al.  (\cite{per77}) suggested
that  all  the  non-evolved  stars  were sitting  on  the  same ZAMS,
whatever   their  metallicity  was.   They  showed  that  this  behaviour  was
reproduced  by  theoretical  stellar  models  if  helium  and metallicity were
related  through a value of \dydz constant and equal to 5.  Recently Fernandes
et  al.  (\cite{fer96}) claimed that stars do not all lie on the same
ZAMS and
measured the observational lower main sequence width in the solar 
neighbourhood. They showed that if the width is entirely due to
chemical composition dispersion in the solar neighbourhood, then
\dydz is greater than 2 in the  corresponding  stars.  

More recently  H{\o}g  et~al.   (\cite{hog97}) and Pagel \& Portinari (\cite{pap98}) 
 have  found \dydz= 3 $\pm$ 2 from a
sample of nearby stars with Hipparcos parallaxes.

 We recall further constraints on these points. 

\subsubsection{The Sun}

The Sun is a milestone in internal structure theory, because its age and
fundamental parameters are known with great accuracy, and helioseismology
(P\'erez  Hern\'andez  \&  Christensen-Dalsgaard \cite{per94})
has brought additional constraints.
All observations have to be reproduced by the theoretical solar model,
with underlying input physics and free parameters.
The \emph{initial}
helium  abundance is a free parameter of the solar model and
is estimated with the luminosity constraint (Christensen-Dalsgaard
\cite{chr82}).
The difference  between  the  present helium value derived
from seismology(Y$\approx$ 0.25) and the initial value  obtained from
calibration (Y $\approx$ 0.275) provides a constraint on the input physics of the model:
it can be explained by invoking the microscopic  diffusion of helium
and heavy elements which has taken
place during  the  evolution  of  the  Sun
(see e.g. Cox et~al.  \cite{cox89}, Richard et al. \cite{RVCD96}, Brun et al. \cite{BTM98}). 
Furthermore the
mixing-length parameter, which enters the  mixing-length
phenomenology of convection is fixed by the radius constraint at a 
value $\approx$ 1.7-1.8.
 
\subsubsection{Binary stars and clusters}

The study of a few visual binary systems of known mass, effective
temperature, luminosity and metallicity provides further
information on low mass stars. The two components of a binary
system, which are assumed to have the same age and initial
chemical composition,
can be modelled simultaneously providing the age of the system, its
initial helium content and the mixing-length ratio (Noels et
al. \cite{noe91}). The method was first applied to the $\alpha$
Centauri system by Noels et al. (\cite{noe91}).
Recently,  Fernandes  et  al.   (\cite{fer98})
studied four Population  I  low  mass  binary systems with
high quality data and determined
the initial helium abundance by mass for three of them with
a precision of 0.02.  A global conclusion of these papers
is that the mixing-length parameter $\alpha $ seems independent
of metallicity, and that a \dydz ratio of 2.3 $\pm 1.5$ 
is appropriate (their Fig. 5).  

Using the high quality observations of Hipparcos
together  with ground-based spectroscopic data, Perryman et al.
(\cite{per98}) and Lebreton  et  al. (\cite{leb97})  determined
the helium abundance by mass, $Y$, of the Hyades with a precision of 0.02
. This abundance does not follow the overabundance of the metallicity 
( +.15 dex). This suggests that a scatter does exist on the \dydz ratio. 
This paper also confirms the universality of the $\alpha$ parameter.

\subsection{Extragalactic observations}

The studies mentioned above have provided the initial helium content of
a few stars of known metallicity which allows to calculate their
helium  to  metal enrichment ratio
\hbox{\dydz = ($Y$ - \Yp)/$Z$}. The primordial
helium abundance $Y_p$ has  been
determined by several groups: for instance, Balbes  et~al. (\cite{bal93}) gave
\Yp=0.227$\pm$0.006 while, more recently, Izotov et al.
(\cite{izo97}) found a higher value \Yp=0.243$\pm$0.003. However, Hogan et al.
(\cite{HOS97}) consider this value as an upper limit. This gives two secure points
of the Y(Z) relationship: $Y_p \approx 0.24$ at $Z=0.$, and $Y_{\odot }=0.275$ at
$Z= Z_{\odot}$. The observations of blue compact galaxies and other systems
have allowed to determine  a \dydz ratio : a recent discussion is given by Izotov et al.
(\cite{izo97}). The value is  2.3 $\pm$ 1.0,  considerably below
former determinations $\approx 4.0$.

\subsection{Nucleosynthesis}   

 On  the other hand
nucleosynthetic predictions  integrated over the whole stellar mass
range lead to \dydz-values ranging  from  about 1 to about 2 depending
on the inclusion of stellar winds; \dydz  can  even reach values of
about 5 if black holes are considered (Maeder \cite{mae92}).
Although helium is expected to increase with metallicity, it must be noted
that metals are only produced by SNe, whereas helium is also produced by mass-loss
of medium-mass stars, in their post-AGB phase.

This body  of results indicates that
a significant dispersion of the helium abundance around the solar-scaled value 
cannot be excluded.

\section{Observational determination of the stellar fundamental parameters}
\label{observations}

\begin{table*}

\caption[ ]{Observational parameters for Sample1. Parallax $\pi$, and
relative error on it $\sigma_{\pi}/\pi$   come from the Hipparcos
main catalogue. The apparent magnitude $V$ is from the Hipparcos Input Catalogue.
$M_{\rm bol}$ and $T_{\rm eff}$ were derived from the bolometric fluxes of
Alonso et al. (\cite{alo95}, \cite{alo96a}).Note that the zero-point of bolometric
magnitudes is not that used by Alonso, but is such that $M_{bol}$ =4.75 for the Sun.
{\rm [Fe/H]} values are a weighted average of individual values from
spectroscopic analyses taken in Cayrel de Strobel et al. (\cite{cayg97}). A quality index "qlt"
was attributed to the  adopted value of [Fe/H] according to the following code:

4: average of at least six determinations obtained with recent high S/N spectra \\
3: average of at least three determinations obtained with recent high S/N spectra \\
2: at least one determination obtained with recent high S/N spectra \\
1: [Fe/H] based on photographic spectra, keeping only high quality work, shown to
be exempt of large systematic errors (often from Hernshaw, see Fuhrmann \cite{F98})   
}

\label{Sample1}

\begin{flushleft}

\begin{tabular}{llllllllllll}

\hline\noalign{\smallskip}

{\rm ~HIC} &  {\rm ~~HD} & ~~$\pi$  &  ~$\sigma_{\pi}/\pi$  & ~$V$ &
$T_{\rm eff}$ & $\sigma_{T_{\rm eff}}$ & $M_{\rm bol}$ &
$\sigma_{M_{\rm bol}}$ & {\rm [Fe/H]}  & qlt &  \\

\noalign{\smallskip}
\hline\noalign{\smallskip}

   171 &  224930A &  80.63 & 0.038 & 5.342 & 5562. & ~80.& 5.220 & 0.087 & -0.76 & 2 & 85 Peg      \\
  3821 &   4614   & 167.99 & 0.004 & 4.576 & 5817. & ~97.& 4.506 & 0.031 & -0.31 & 3 & $\eta$ Cas  \\
  5336 &   6582   & 132.40 & 0.005 & 5.789 & 5339. & ~82.& 5.619 & 0.032 & -0.76 & 2 & $\mu$~Cas   \\
  7918 &  10307   &  79.09 & 0.010 & 4.451 & 5874. & ~99.& 4.401 & 0.038 & -0.02 & 3 &            \\
  7981 &  10476   & 133.91 & 0.007 & 5.874 & 5172. & ~78.& 5.694 & 0.033 & -0.20 & 2 &            \\
  8102 &  10700   & 274.17 & 0.003 & 5.680 & 5388. & ~81.& 5.550 & 0.031 & -0.56 & 4 & $\tau$ Cet  \\
 10629 &  13783   &  25.82 & 0.041 & 5.360 & 5501. & ~86.& 5.250 & 0.095 & -0.55 & 1 &            \\
 10644 &  13974   &  92.20 & 0.009 & 4.684 & 5591. & ~67.& 4.594 & 0.036 & -0.30 & 1 &            \\
 13402 &  17925   &  96.33 & 0.008 & 6.040 & 5123. & ~75.& 5.757 & 0.037 & ~0.10 & 2 &            \\
 16537 &  22049   & 310.75 & 0.003 & 3.726 & 5076. & ~86.& 5.982 & 0.030 & -0.17 & 3 & $\epsilon$ Eri\\
 17147 &  22879   &  41.07 & 0.021 & 4.758 & 5798. & ~87.& 4.658 & 0.054 & -0.88 & 4 &            \\
 26779 &  37394   &  81.69 & 0.010 & 5.781 & 5185. & ~53.& 5.611 & 0.037 & -0.20 & 1 &            \\
 27072 &  38393A  & 111.49 & 0.005 & 3.826 & 6260. & 104.& 3.806 & 0.032 & -0.09 & 3 & $\gamma$ Lep A\\
 36640 &  59984A  &  33.40 & 0.028 & 3.519 & 5928. & 101.& 3.429 & 0.068 & -0.81 & 4 &            \\
 39157 &  65583   &  59.52 & 0.013 & 5.863 & 5242. & ~59.& 5.703 & 0.041 & -0.61 & 1 &            \\
 44075 &  76932   &  46.90 & 0.021 & 4.176 & 5727. & ~77.& 4.106 & 0.054 & -0.97 & 4 &            \\
 50139 &  88725   &  27.67 & 0.037 & 4.960 & 5669. & ~66.& 4.830 & 0.085 & -0.65 & - &            \\
 56997 & 101501   & 104.81 & 0.007 & 5.422 & 5342. & ~91.& 5.312 & 0.034 & ~0.03 & 1 &            \\
 61317 & 109358   & 119.46 & 0.007 & 4.646 & 5867. & ~99.& 4.586 & 0.034 & -0.08 & 3 &            \\
 64394 & 114710   & 109.23 & 0.007 & 4.452 & 5964. & ~93.& 4.412 & 0.033 & ~0.04 & 4 &            \\
 70319 & 126053   &  56.82 & 0.018 & 5.043 & 5646. & ~68.& 4.953 & 0.050 & -0.20 & 2 &            \\
 73005 & 132142   &  41.83 & 0.015 & 5.877 & 5098. & ~67.& 5.697 & 0.044 & -0.55 & 1 &            \\
 73184 & 131977   & 169.32 & 0.010 & 6.864 & 4605. & ~43.& 6.424 & 0.037 & ~0.01 & 1 &            \\
 80837 & 148816   &  24.34 & 0.037 & 4.212 & 5851. & ~62.& 4.132 & 0.086 & -0.80 & 3 &            \\
 84478 & 156026   & 167.56 & 0.006 & 7.451 & 4345. & ~39.& 6.881 & 0.033 & -0.34 & 2 &            \\
 88972 & 166620   &  90.11 & 0.006 & 6.174 & 4947. & ~74.& 5.914 & 0.033 & -0.20 & 1 &            \\
 94931 & +41 3306 &  28.28 & 0.030 & 6.097 & 5004. & 139.& 5.867 & 0.073 & -0.87 & 2 &            \\
 96100 & 185144   & 173.41 & 0.003 & 5.875 & 5227. & ~89.& 5.705 & 0.031 & -0.23 & 1 &            \\
 96895 & 186408A  &  46.25 & 0.011 & 4.286 & 5763. & ~90.& 4.226 & 0.038 & ~0.06 & 2 & 16 Cyg A     \\
 96901 & 186427B  &  46.70 & 0.011 & 4.567 & 5767. & ~90.& 4.487 & 0.039 & ~0.06 & 2 & 16 Cyg B     \\
 98792 & 190404   &  64.17 & 0.013 & 6.317 & 5001. & ~75.& 6.067 & 0.042 & -0.15 & 1 &            \\
104214 & 201091A  & 287.13 & 0.005 & 7.490 & 4323. & ~50.& 6.890 & 0.032 & -0.05 & 1 & 61 Cyg A     \\
114622 & 219134   & 153.24 & 0.004 & 6.497 & 4785. & ~59.& 6.167 & 0.031 & ~0.00 & 1 &            \\

\noalign{\smallskip}

\hline

\end{tabular}

\end{flushleft}

\end{table*}

\begin{figure}
\resizebox{\hsize}{!}{\includegraphics{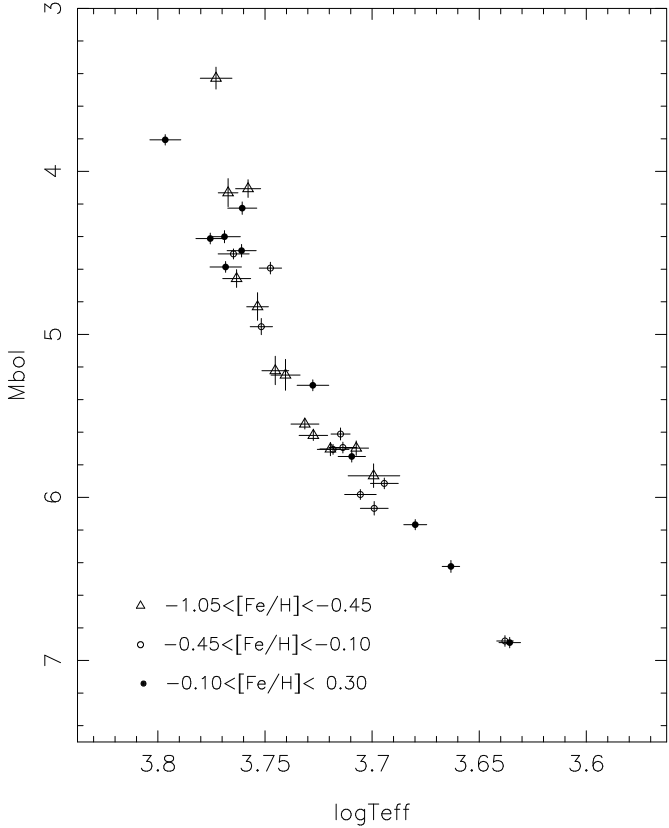}}

\caption{The  Hipparcos  HR  diagram of the 33 best known nearby stars (Sample
1,  see  text).   The  accuracy  on  the parallax is better than 5
per cent and
bolometric  fluxes are available for each star.  Individual errors bars are 
given  for  each  object  by  the  horizontal and vertical sizes of the cross.
Filled  circles,  open  circles  and open triangles respectively correspond to
stars  with  [Fe/H]  values  in the interval [-0.10, 0.30], [-0.45, -0.10] and
[-1.05,-0.45]}

\label{Fig1}
\end{figure}

We study an homogeneous sample of 114 late-type nearby stars of spectral types
in the range from F to late K, carefully selected by M.-N.  Perrin. They are part
of the proposal 132 (M.-N. Perrin), accepted as an Hipparcos program in 1982.
Later on, the proposal was updated as the INCA proposal 011 by A. Baglin,
M.-N. Perrin, Y. Lebreton and R. and G. Cayrel. 
These  stars  are  closer  than  about  25  parsecs which ensures an excellent
accuracy  of their parallax determination by Hipparcos.

Among  these, we have retained the stars which have been submitted
to  detailed  spectroscopic  analysis  from the ground and which appear in the
last version   of  the {\it   Catalogue of [Fe/H] determinations: 1996 edition} by   
G. Cayrel de Strobel et al. (\cite{cayg97}).  Their metal to hydrogen ratio [Fe/H],
i.e.  the logarithm of
the iron to hydrogen ratio (by number of atoms) relative to the solar value, ranges
from  about  -1.0  to  about +0.3 dex corresponding to Population I and thick-disk
population.   For each star we assign an averaged abundance 
determined from the above catalogue (for more details see the caption of table 1).
  The mean internal error on [Fe/H] is of the
order of 0.1 dex . This does not include systematic errors, as the fact that all abundances 
in the catalogue are not corrected for NLTE effects. In determinations based on spectra taken with
solid state detectors, with high quantum efficiency, the random error coming from equivalent
width measurements enters for about 0.03 dex only, the remaining part coming from  uncertainties
in the fundamental parameters \Teff and $ \log g$.  

We  eliminated the  suspected  unresolved binaries and 
only  kept stars with parallaxes determined with an accuracy better than 5 per
cent. Then, among the remaining stars,  we extracted three
homogeneous and independent subsamples.

Sample  1  is  constituted of 33 stars with directly determined bolometric
fluxes,  so  the  problem  of  the  value  of  the  bolometric  correction  is
eliminated.  The relevant data for Sample 1 are listed in Table~\ref{Sample1}.
The  fluxes  were derived by Alonso et~al.  (\cite{alo95}) with an accuracy of
about  2  per  cent  by integrating UBVRIJHK photometry.  The average absolute
error  on bolometric magnitude resulting from  the parallax and bolometric
flux uncertainty is of about 0.03 to 0.06 magnitude.  The effective temperature was
obtained by Alonso et~al.  (\cite{alo96a}) from the bolometric flux by the 
Infra-Red Flux Method (IRFM) (Blackwell \& Shallis \cite{BS77}). The method makes 
use of a grid of theoretical model   line-blanketed  flux  distributions  (Kurucz
\cite{kur91}).  The mean internal error on effective temperature is about 80 K
for   temperatures   greater   than  4200  K  (Alonso  et~al. \cite{alo96a}).
Fig.~\ref{Fig1}  shows the resulting positions of the stars in the (log \Teff, 
\Mbol) diagram with their individual error bars.
The striking feature of Fig.~\ref{Fig1} is the lack of clear separation between stars
having a solar metallicity, and those in the metallicity range [-1.0;-0.5],
confirming with high quality data the result already obtained by Perrin et al.
(\cite{per77}), on less accurate data. 
In Fig.~\ref{Fig2ab} the sample of stars plotted in Fig.~\ref{Fig1} is
 split into 2 subsamples corresponding 1) to stars of solar metallicity,
plotted in Fig.~2a and
2) to moderately deficient stars, plotted in Fig.~2b.  Theoretical
isochrones are also plotted in Fig.~\ref{Fig2ab} but this will be
discussed later in Sect.~\ref{results}.

Sample  2  is constituted of 64 stars with effective temperatures derived from
detailed  spectroscopic  analysis.  For each star at least one determination
of the effective temperature and of [Fe/H] was made with modern
detectors  (CCD  or  Reticon).   We  adopt  a  mean  internal uncertainty
on effective
temperature  of  150  K.  Bolometric magnitudes were obtained using bolometric
corrections   from  Alonso  et~al.   (\cite{alo96a},  \cite{alo96b}),  with  a
correction  of  the  zero  point giving a bolometric magnitude of 4.75 to the
Sun.   The  resulting positions of the stars in the (log \Teff, \Mbol) diagram
are   plotted   in   Fig.~\ref{Fig3ab}   and   will   be  discussed  later  in
Sect.~\ref{results}.

Sample  3  is constituted of 15 stars with effective temperatures derived from
the  ($V-K$) color index and from [Fe/H] according to the results from Carney
et~al. (\cite{CLLA94}), Johnson et~al. (\cite{JMIW66}),  Koorneef (\cite{Koo83}).
The mean internal uncertainty on effective
temperature  is of about 75 K.  The bolometric magnitudes are derived from the
$V$  magnitude  given  in  the  Hipparcos input catalogue, using bolometric
corrections  from Alonso et~al.  (\cite{alo96a}, \cite{alo96b}), with the same
 zero-point.  Fig.~\ref{Fig4ab} shows the resulting positions of
the  stars in the (log \Teff, \Mbol) diagram with their error bars and will be
discussed in Sect.~\ref{results}.
Very   clearly   Fig.~\ref{Fig3ab}   has   a   much   larger   scatter  than
Fig.~\ref{Fig1}, 2 and 4,
   showing  that  the  effective
temperatures  derived from  detailed  analyses  are  not tightly connected to the true
effective  temperatures.   The  effective temperatures derived from $(V-K)$ on
the  contrary  are  tightly bound to those obtained by the IRFM,
 as  it  can be directly checked in the full Alonso sample.  It is not
surprising, as effective temperatures from detailed analyses are affected by
other  parameters:  gravity, departures from LTE in ionisation equilibria,
and  a  variety  of  different  techniques  in fixing the triad (\Teff, log g
and [Fe/H]).

\section{Theoretical stellar models}
\label{models}

\subsection{Input physics}

The   stellar  evolution  calculations  were  computed  with  the  CESAM  code
(Morel \cite{mor97}) in which we have included appropriate input physics 
for the mass range under investigation.

The    CEFF    equation    of    state   (Eggleton   et   al.    \cite{eff73},
Christensen-Dalsgaard \cite{chr91}), was used. It includes the Coulomb corrections to
the  pressure and is appropriate when modeling low-mass stars of mass greater than
about 0.6 \Msun (Lebreton \& D\"appen \cite{leb88}).

We used the nuclear reaction rates given by Caughlan \& Fowler (\cite{cau88}).

We  calculated  the initial composition of the models either from the Grevesse
\&  Noels  (\cite{gre93})  solar mixture (GN93 mixture) or from a GN93 mixture
where  the  $\alpha$-elements O, Mg, Si, S, Ca, Ti are enriched relative to
the  Sun  ([$\alpha$/Fe]=  +0.4  dex). Such an enrichment of $\alpha$-elements 
is
observed  in  metal  deficient stars with metallicities [Fe/H] lower than -0.5
(Wheeler et~al.  \cite{whe89}, Fuhrmann \cite{F98}) and its effect on models
has to be taken into account.

We  took  the  most  recent  OPAL  opacities (Iglesias \& Rogers \cite{igl96})
provided  for  the  two  mixtures  considered.   They were complemented at low
temperatures  by  atomic  and  molecular opacities from Alexander \& Ferguson
(\cite{ale94})  for  the  GN93  mixture  or from Kurucz (\cite{kur91}) for the
$\alpha$-enriched  mixture. Low and high temperatures tables were fitted
carefully at a temperature of about 10~000~K, depending upon the chemical composition.

In most models the atmosphere was described with an Eddington's $ T(\tau)$ law
which  is  easy  to use.  In
order  to estimate the uncertainties resulting from this choice we built a few
models  using  $T(\tau)$ laws derived from the ATLAS9 atmosphere models (Kurucz
\cite{kur91}).   This  requires  to  calculate detailed atmosphere models with
ATLAS9 for many values of the gravity, of the effective temperature and of the
metallicity   and   then   to   derive   the   corresponding  $T(\tau)$  laws
(see Morel et al. \cite{MVP94}).  Interpolation of the $T(\tau)$ laws is then 
performed to model
the  considered  star.  
We checked on a test-case, that
the use of those better atmospheres does not shift the position of the model in the HR 
diagram by  an amount larger than the observational error bars.

Convection was treated according to the mixing-length theory of B\"ohm-Vitense
(\cite{boh58}).   The mixing-length parameter $\alpha$, ratio of mixing-length
to  pressure  scale-height,  is  a free parameter of the models. As shown by
Fernandes et~al.  (\cite{fer98}) the  Sun and four visual binary systems
spanning a wide range of masses and  metallicities could be calibrated with
very close values of $\alpha$. Moreover the slope of the Hyades
main-sequence is well-reproduced with a solar $\alpha$-value
(Perryman et al., \cite{per98}). On the other hand, Ludwig et~al.
(\cite{lud99})  made a calibration of the mixing-length based on 2-D radiation
hydrodynamics  simulations  of  solar-type  surface  convection
and found that variations  of  $\alpha$ of about 10 per cent around the solar
value  are  expected  in  the  domain  of  effective  temperature, gravity and
metallicity  studied here.  
We  adopt  the solar mixing-length value in our  calculations. For 
unevolved stars a change of $\ell/H_{\rm p}$ of $\pm$ 0.15 produces a change 
in  effective temperature of the order of 50 K,  
smaller than the mean observational error on $T_{\rm eff}$. 

With  the  input  physics  described  above  the calibration in luminosity and
radius  of  a  solar  model  having  $(Z/X)_{\sun}$~=0.0245,   where  $X$ is the
hydrogen  abundance  by  mass  (Grevesse  and  Noels  \cite{gre93}) requires a
mixing-length  $\ell=  1.64~H_{\rm  p}$, an initial helium abundance $Y$=0.266
and  a  metallicity  $Z_{\sun}$=0.0175.   

\subsection{Grids of stellar models and isochrones}

In order to determine the ZAMS position , we calculated zero and terminal
age main sequences (ZAMS and TAMS)
for 10 mass values ranging
from 0.5 to 1.4 \Msun, 5 helium values ranging from 0.18 to
0.43 and metallicities values $Z$= 0.004, 0.007,
0.01,  0.015, 0.02, 0.03, 0.04 and 0.06.

To discuss the global features of the HR diagram and the position
of particular objects, we calculated detailed evolutionary
sequences from the ZAMS to the beginning of the red-giant branch,
for 14 masses ranging from 0.5 to 5 \Msun and we derived isochrones
using the Geneva isochrone program (Maeder \cite{mae74}). We
chose metallicities corresponding to  the  observational  range  (i.e.
[Fe/H]=+0.3,  0.0,  -0.5  and  -1.0)  and  solar-scaled values  of the helium
abundance given by :

$$Y(Z) = Y_{\rm p} + (Z/Z_{\odot}) \,(Y_{\odot}-Y_{\rm p})$$

\noindent with $Y_{\rm p}=0.227$ (Balbes et al. \cite{bal93}). This implies a  
\dydz value of 2.2.
 
  Two distinct grids of models were calculated
in  the  metal  deficient  cases ([Fe/H] = -0.5 and -1.0):  a grid with normal
solar mixture and a grid with an $\alpha$-elements enhancement of 0.4 dex.
For the solar mixture grid the
metallicity  $Z$  in mass fraction is related to [Fe/H] by:

$$ {\rm [Fe/H]}  = \log(Z/X) - \log{(Z/X)}_{\sun}$$

\noindent where $(Z/X)_{\sun}$ is the ratio of the solar mixture considered.

For the $\alpha $ -elements enhanced mixture the relation becomes:

$$ {\rm [Fe/H]}  =  \log(Z/X) - \log{(Z/X)}_{\sun} -0.305$$

\noindent where $ 0.305 =   \log{(Z/X)}_{\sun, \alpha}-\log{(Z/X)}_{\sun} $,
difference between a solar $\alpha $-elements enhanced
mixture and a `normal' solar mixture.

The  detailed results of all the stellar models computed for that work will be
published  separately   (Lebreton et~al., in preparation) and  are
available on request.

\begin{figure*}[htbp]
\centerline{
\resizebox{\hsize}{!}{\includegraphics{8691.f2}}
}
\caption{Same  as  Fig.~\ref{Fig1}  (i.e.   stars  of Sample 1) but split 
in two metallicity domains: stars of solar metallicity and close to it (Fig.~2a)
and moderately deficient stars (Fig.~2b). Theoretical isochrones are
overlaid on the observational data. Fig.~2a: the lower isochrone  (8 Gyr) is
for [Fe/H]=$-0.5$, $Y$=0.256 and [$\alpha $/Fe]=0.4; the upper isochrone
 (10 Gyr) is for [Fe/H]=0.3, $Y$=0.32 and [$\alpha$/Fe]=0.0. The dashed line
 is a solar ZAMS. The young star $\gamma $~Lep is the brightest star in the
 figure. Fig.~2b: the lower isochrone (10 Gyr) is for [Fe/H]=$-1.0$, $Y$=0.236
 and [$\alpha $/Fe]=0.4; the upper isochrone (10 Gyr) is for [Fe/H]=$-0.5$,
 $Y$=0.256, [$\alpha $/Fe]=0.4. All stars, but one, are lying above the
 lane defined by these two isochrones. 
}

\label{Fig2ab}

\end{figure*}
\begin{figure*}[htbp]

\resizebox{\hsize}{!}{\includegraphics{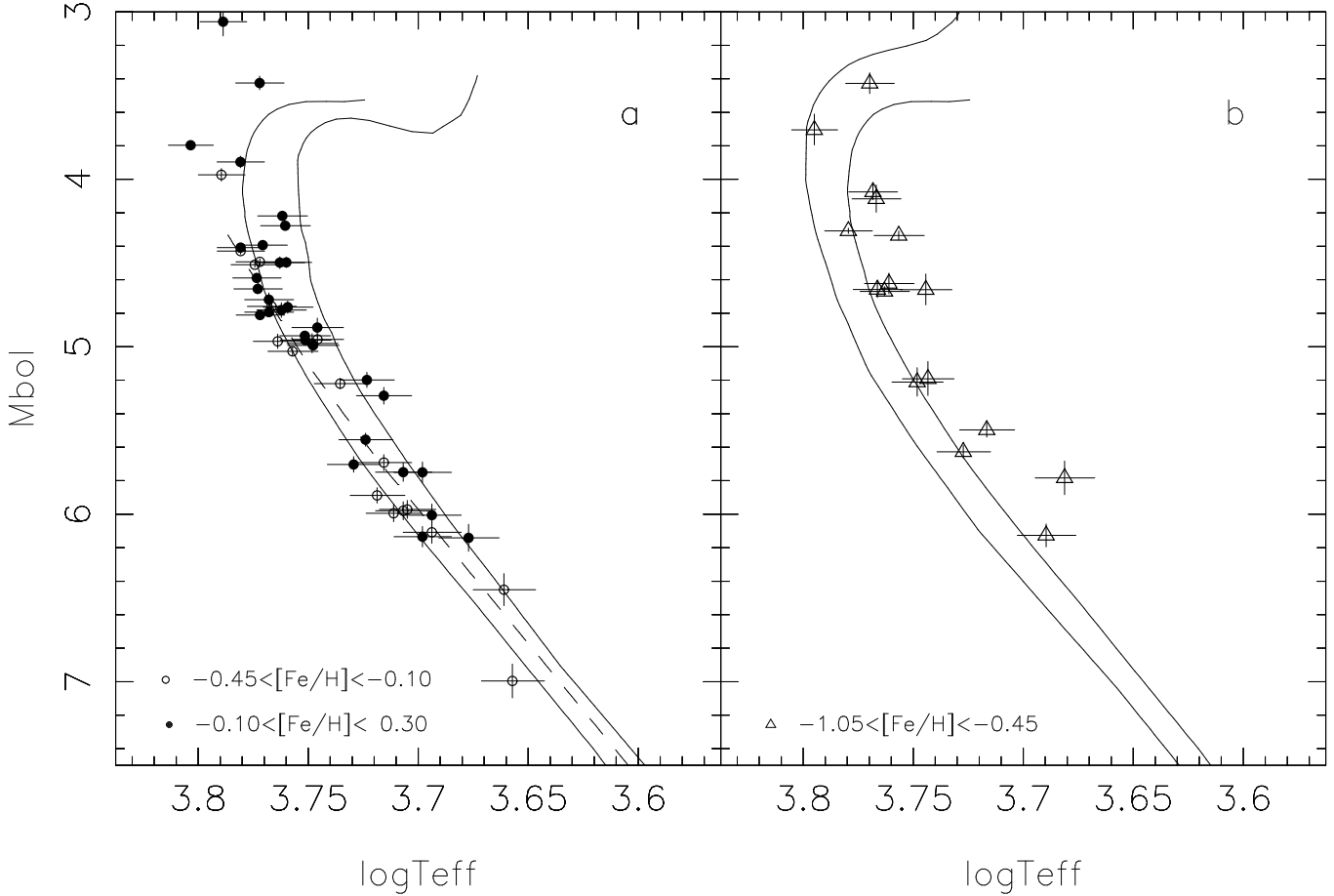}}

\caption{Same  as  Fig.~\ref{Fig2ab}  for  stars of Sample 2 ($M_{\rm bol}$ is
obtained  from the Hipparcos parallax and $V$ magnitude and $T_{\rm eff}$ from
detailed spectroscopic analyses, see Sect.~\ref{observations})}

\label{Fig3ab}

\end{figure*}
\begin{figure*}[htbp]

\resizebox{\hsize}{!}{\includegraphics{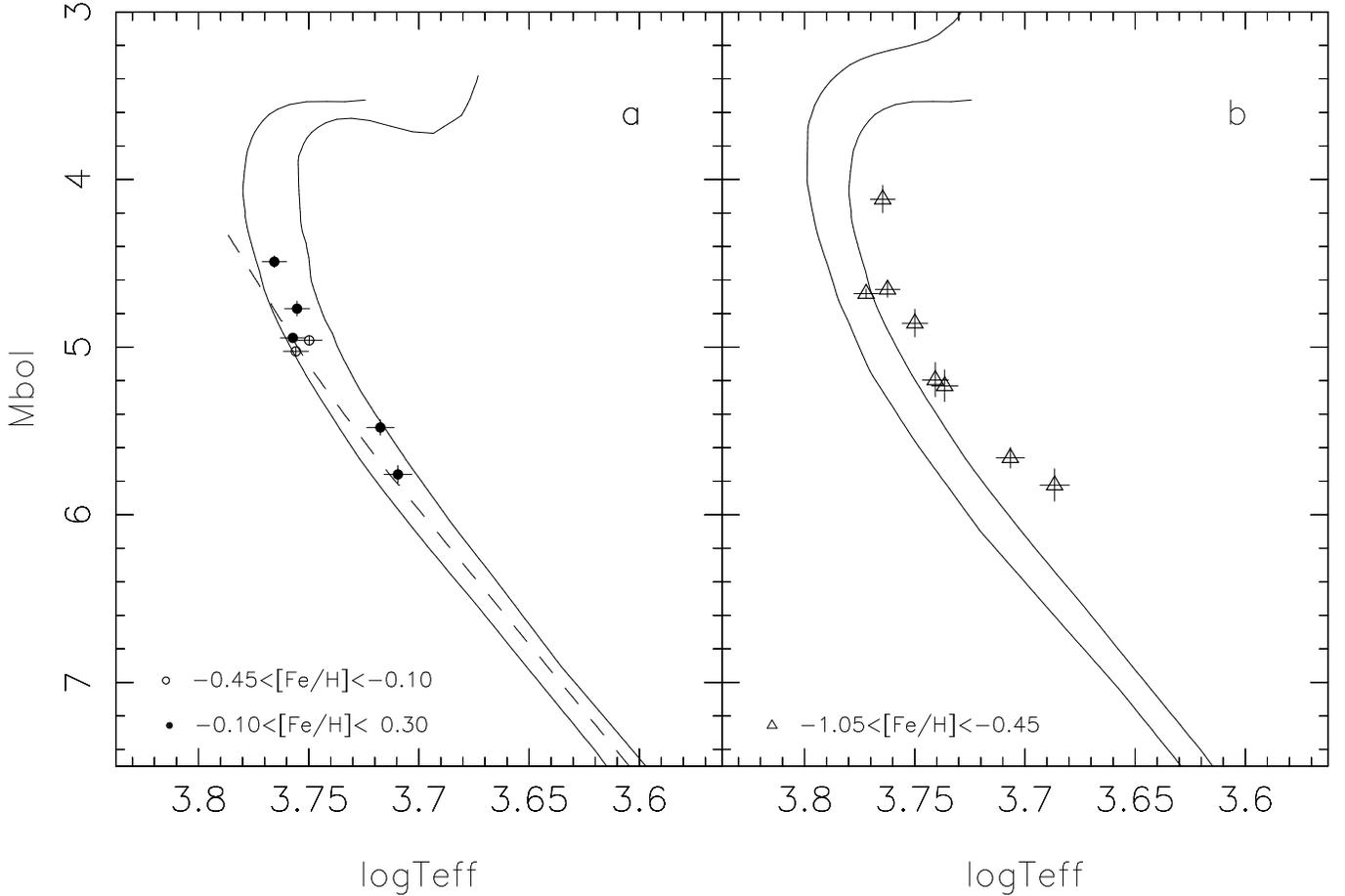}}

\caption{Same  as  Fig.~\ref{Fig2ab}  for  stars of Sample 3 ($T_{\rm eff}$ is
obtained    from    the    (V-K)    color   index   and   from   [Fe/H],   see
Sect.~\ref{observations})}

\label{Fig4ab}

\end{figure*}

\section{Comparison between theoretical isochrones and observational data}

\label{results}

We  present in Fig.~2a and in Fig.~2b the HR diagram of the 33 stars of Sample
1.   We  have split the sample in two subsamples:  Fig.~2a corresponds
to   stars   of   solar   metallicity   or   close   to  it  ($-0.45~\leq~{\rm
[Fe/H]}~<~+0.30$)  while  Fig.~2b is constituted of the moderately metal
deficient  stars ($-1.05~\leq~{\rm [Fe/H]}~<~-0.45$).  We have superimposed on
those  diagrams  theoretical  isochrones  associated  to  the  limits  of  the
metallicity  ranges considered.  Isochrones with [Fe/H] values of -0.5 and -1.0
have  been  derived  from models calculated with an $\alpha$-elements enriched
mixture  and  have  solar-scaled helium  values, $Y$=0.256 and 0.236 respectively.
The metal-rich isochrone has a solar $\alpha$-elements ratio and $Y$=0.32.
We have also  plotted  the position of the solar ZAMS on Fig.~2a.  Isochrones
are  given  for  ages  in  the range 8-10~Gyr representative of the age of the
galactic disk.

Fig~2a  shows that the stars of solar metallicity or close to it lie
in  the  region  defined  by  the  theoretical  models  corresponding to their
associated  metallicity  range.  We notice the particular position of $\gamma$
Lep (HIC 27072) which is a very young star lying close to the ZAMS.

However, it is  worth  to  note  that the slope of the theoretical main sequence
agrees  well with the observational slope.  This, once more, is in favor of the uniqueness
of the mixing-length parameter in low-mass stars. Furthermore
the width of the global observed main-sequence band for metal-rich stars, which is around 0.3 mag,
and the theoretical width corresponding to the same metallicity
range with solar-scaled helium abundances are quite similar
(see also Fig~\ref{Fig6}). 

Turning  now to the moderately deficient stars plotted in Fig.~2b \emph{we note that
all the stars but one are outside the theoretical band corresponding to their metallicity}.
Stars are located on theoretical isochrones with a metallicity higher than
their observed metallicity.

The  same  behaviour  is  seen  in the two other selected samples in which the
effective   temperatures  of  stars  have  been  determined  independently  by
different  authors (see Sect.~\ref{observations}).  Fig.~3a and 3b are similar
to  Fig.~2a  and  2b  but for the 64 stars with effective temperatures derived
from detailed spectroscopic analysis (Sample 2) while Fig.~4a and 4b represent
the  situation  of  the  15  stars  of  Sample  3.   Although the error bars on
effective  temperature are greater in Fig.~3a and 3b and although they are few
objects in Fig.~4a and 4b, the same tendency is found,
strengthening the conclusion that classical
theoretical  isochrones and actual observations of moderately
metal deficient stars do not match.

Note that, in the confrontation  of observational  position of stars
with stellar isochrones,
errors on [Fe/H] and  [$\alpha$/Fe] which are in the range 0.10-0.15 dex, are
responsible for a hidden enlargment of the error boxes, because the chemical 
composition of the models to be compared with the observed stars are not
known exactly.

\section{Resolving the metal-poor discrepancy}
 
The most natural attempt to solve the discrepancy is by trying to adjust the ``free''
parameters, the helium content Y being the first target. However, for the stars in the
metallicity bin [$-1.05,-0.45$], a ${\rm Y} \leq 0.2$ would be required, a 
cosmologically unacceptable value. A change in  $\alpha=\ell/{\rm H_p}$ does
not help either, as it
  is producing a negligible effect
in the low part of the main sequence under consideration here.

So, as noted in Lebreton et~al. (\cite{LPFCCB97}) the discrepancy in the expected and
actual positions of metal-poor stars needs either a drastic change in the
zero-point of effective temperature (of the order of 200 K) or accepting the
view that  other processes must be included in the interpretation of the
observations.
 It would be very
difficult to plead a large correction in the effective temperatures at
metallicity ${\rm [Fe/H]}=-0.7$ and none at solar metallicity.
So, we do not further consider this explanation and, now focus on
the second one.

 Two processes are
considered in this section: (i) departures to LTE in the determination of [Fe/H]
(ii) microscopic diffusion of helium and other heavy elements. 

\subsection{non-LTE departures for iron}

All the ${\rm [Fe/H]}$ values we have used were computed with the LTE 
approximation. Computing NLTE abundances for iron is a formidable task. However,
it must be noted that Bikmaev et~al. (\cite{BB90}) have claimed that a
substantial overionisation occurs for Fe~I in subdwarfs, and only \hbox{Fe II}
lines should be used for determining iron abundances. Holweger and coworkers 
(private communication, 1993) found the same phenomenon in trying to compute
departures from LTE for the very metal-poor star ${\rm CD}-38^\circ 245$. More 
recently, Nissen et~al. (\cite{nis97}) found that spectroscopic gravities in
subdwarfs were systematically lower that the gravities derived from Hipparcos
distances. Finally, Th\'evenin \& Idiart (\cite{TI99}) have undertaken NLTE 
computations of iron abundances in one hundred metal-poor stars, and found
non-LTE corrections for several of our stars, of the order of 0.15 dex, for the
mean metallicity (${\rm[Fe/H]}=-0.72$) of our  $-1.05 ~\leq~{\rm [Fe/H]}~<~-0.45 $ subsample.
Non-LTE corrections are negligible for stars with solar metallicities
(same reference). The metallicity of isochrones to be compared to the set of stars
with a mean LTE metallicity [Fe/H]=-0.72 is therefore the true NLTE metallicity
[Fe/H]=-0.57. 
The situation is improved but a clear departure 
is still existing (Fig.~\ref{Fig5}).

\subsection{Sedimentation of heavy elements}
 
The effects of sedimentation of heavy elements have recently been computed
by  Morel \& Baglin (\cite{MB99}), for stars of metallicity
${\rm [Fe/H]}=-0.7$, $-0.9$, $-1.2$,$-1.7$, and for 5 stellar masses, 0.85, 0.80, 0.75,
0.70, and 0.60 $M_{\odot}$. To produce significant effects in the HR
diagram microscopic diffusion has to proceed on a sufficiently long
timescale. This condition is fulfilled for the thick disk subsample
under consideration here: the 10 Gyr isochrones of the Morel \& Baglin paper
are a  suitable choice for our  subsample. Two effects are present:
 
After a time of 10 Gyr, the stratification of the heavy elements translates the 
evolution point
in both $\log T_{eff}$ and $M_{bol}$.
 This effect is small: at 0.7 $M_{\odot}$
and an initial metallicity of ${\rm [Fe/H]}=-0.75$ the point is moved by
$\Delta M_{\rm bol}=-0.029$ and $\Delta \log T_{\rm eff}
=-0.0004$. But the surface
metallicity of the star has dropped to ${\rm [Fe/H]}=-0.81$. Therefore, the
theoretical position of a star of present metallicity ${\rm [Fe/H]}=-0.75$, must
be computed with a modified initial composition, leading after a 10 Gyr evolution,
to the metallicity we observe now. This is the largest part of the correction, called
the calibration correction by the authors. From their table~2, and their Fig. 5
we have computed the corrections to be applied to a standard isochrone for the
mean metallicity  (${\rm[Fe/H]}=-0.72$)
of our sample in the range $-1.05 ~\leq ~{\rm [Fe/H]}~<~-0.45 $.

\subsection {Cumulated effect}

\begin{figure}[]
\resizebox{\hsize}{!}{\includegraphics{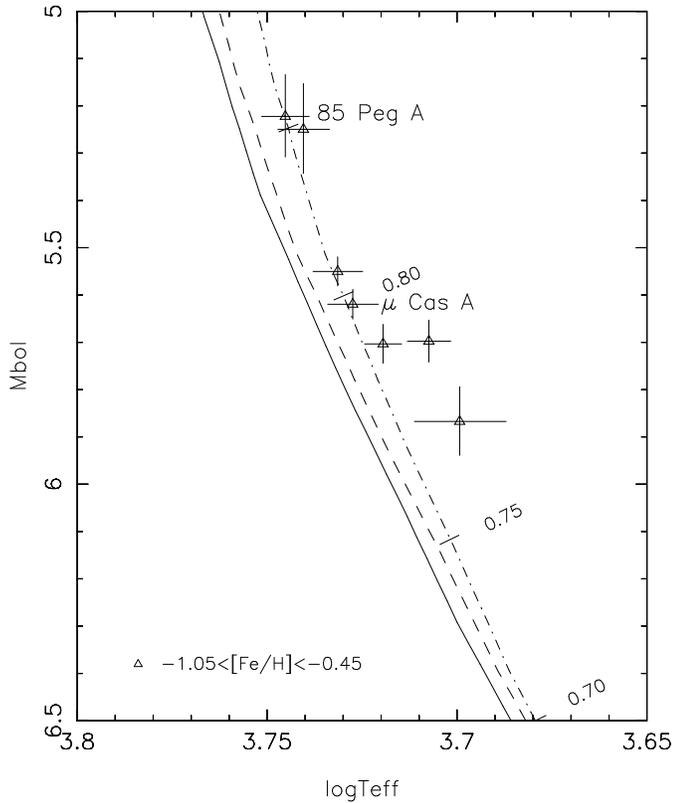}}
\caption{ This figure  is a large scale HR diagram for the subsample of unevolved
stars in the LTE metallicity range -1.05 $<$ [Fe/H] $<$ -0.45. The full line is 
a standard isochrone for the mean LTE metallicity of the sample [Fe/H] = -0.72. The
dashed line is the standard isochrone corresponding to the NLTE mean metallicity
of the sample [Fe/H] = -0.57. The dot-dashed line corresponds to an isochrone 
including microscopic diffusion of the elements for an age of 10 Gyr. The surface
metallicity is the NLTE value [Fe/H]=-0.57, but the initial metallicity, which is 
very close to the mean interior metallicity was [Fe/H] $\approx -0.5$. The fit is 
now satisfactory. The tick marks show the value of the mass along the upper
isochrone. The tick mark near 85 Peg A corresponds to 0.85 M$_{\odot}$ 
 (not labeled for clarity)}
\label{Fig5}
\end{figure}

Applying these corrections to the dashed curve in
Fig.~\ref{Fig5} produces the dot-dashed isochrone. No systematic departure appears any more
between the observations and the theoretical isochrone corresponding to the present
surface metallicity of the stars. 
The initial discrepancy noted at the beginning of this section is then completely
removed. The only star presenting a $2\sigma$ discrepancy is the star HD~132142 at (3.707,
5.697). But
there is a single spectroscopic determination of its metallicity and its photometry 
indicates a solar metallicity. So this exception is not a serious worrying.

\section {Binary stars}

One more constraint can be obtained from the stars of the sample which are 
component of a binary system having known orbit and masses. One of us (JF) has 
already used a few of the best known binaries in order to check the predictions 
of theoretical evolution computations (Fernandes et~al. \cite{fer98}).  It is 
worth to see if, when known, the dynamical mass of the object confirms, or 
disproves, its location along the proper isochrone obtained above. The most
interesting case is the binary $\mu$~Cas, the only metal-poor star, having
a well determined mass.

\subsection {$\mu$~Cas}

Drummond et~al. (\cite{dru95}) have derived the mass of $\mu$~Cas from its orbit,
with the help of speckle interferometry to see its very faint companion
($\Delta m\approx 5.0$). 
The
mass of $\mu$~Cas~A, readjusted with the better distance provided by Hipparcos
is 0.757 $\pm$ 0.06 $M_{\odot}$. The Alonso et~al. (\cite{alo96a}) effective
temperature for (A+B) is 5315 $\pm$ 82 K, which corrected for the presence
of the companion becomes 5339 $\pm$ 85 K, in good agreement with Fuhrmann
(\cite{F98}) effective temperature of 5387 $\pm$ 80 K. We obtain 
$\rm{M_{bol}}= 5.634 \pm 0.035 $ for $\mu$~Cas~A from its parallax, and again 
correcting for the presence of the companion. The position of the star is 
plotted with respect to the uncorrected and corrected isochrones in Fig.~\ref{Fig5}. The
evolutionary masses  along the dot-dashed isochrone are shown with tick-marks.
The location of $\mu$~Cas is near 0.8 $M_{\odot}$, within the limits of
the dynamical mass 0.757 $\pm$ 0.06 $M_{\odot}$. Unfortunately, the error
bar on the mass is still a bit large to add an interesting new constraint.
The influence of the uncertainties on the values of the various parameters of the models
on the position of $\mu$ Cas in the HR diagram are documented in table 2
(see Lebreton \cite{leb98} for more details).

\begin{table}[h] 
\caption[]{budget of the impact of parameter variations on the position
of a star like $\mu $ Cas in the HR diagram. The reference model has an age
of 12 Gyr, a mass of 0.757 $M_{\odot}$, Y = 0.245, [Fe/H] = -0.86, and 
[$\alpha$/Fe]= 0.3 dex} 
\begin{flushleft}
\begin{tabular}{llll}
\hline
 parameter & increment & $\delta \log T_{eff} $ & $\delta M_{bol}$ \\
 \hline
 mass(\Msun) & 0.06 & 0.031  & -0.607 \\
 age (Gyr) & 2 & 0.005 & -0.150   \\
 Y    & -0.015 & -0.012  & 0.172 \\
 $[$Fe/H$]$ (dex)  & -0.15 & -0.015  & 0.188   \\
 $[\alpha$/Fe$]$ (dex)   & 0.10  & -0.004   & 0.064 \\
 $\alpha$ (MLT)  & -0.15 & -0.004 & 0.004 \\
 atmosphere & Eddington/Kurucz & -0.001 & 0.001 \\
 diffusion & yes/no & -0.007 & 0.040 \\
 low-T opacity & +30 per cent & -0.049 & 0.742 \\
 \hline
 \end{tabular}
 \end{flushleft}
 \end{table}

\subsection {85~Peg}
 Fernandes et~al.  (\cite{fer98}) showed that, using the data available 
on  this object, there was no theoretical solution fitting the two components .   
Later on, Martin \& Mignard (\cite{MM98}), have restudied several binaries from Hipparcos
data, and shown that it is
very difficult to escape the conclusion that the mass of the primary and the
secondary are very similar, notwithstanding the fact that the secondary is 3.2
magnitude fainter than the primary. New determinations of the effective temperature of
 85~Peg~A by Th\'evenin \& Idiart (\cite{TI99}) and C. Van't Veer (private communication,
1998) confirms that 85~Peg~A has a metallicity very similar to that of
$\mu$~Cas~A, an effective temperature of 5550 $\pm$ 100 K, and 
$\rm{ M_{bol}}$=5.22 $\pm$ 0.06. Its location in the HR diagram (Fig.~\ref{Fig5})
is pratically on the same isochrone as $\mu$~Cas, and corresponds to a mass
of 0.85 $M_{\odot}$,  higher than the mass given
by Martin \& Mignard (\cite{MM98}) (0.705  $M_{\odot}$), but  below the
mass $0.95\pm 0.1$ given by Duquennoy \& Mayor (\cite{DM91}) from the spectroscopic orbit.
85~Peg~A does not deviate any more from the proper isochrone, once its high [$\alpha $/Fe]
 ratio, its true iron abundance
 corrected for NLTE effects, and sedimentation of heavy elements, are all taken into account.
The problem is with 85~Peg~B,
which is too massive for its absolute magnitude. The colour $(V-K)$ of 85~Peg~B can be derived 
from the V and K magnitudes of (A+B) , V=7.75, K=3.94 (Johnson et al. \cite{JMM68}),
the magnitudes $V_A$= 5.81 and $V_B$= 9.0, and assuming that $(V-K)_A = 1.59 \pm 0.02$ 
from its effective temperature and 
Alonso calibration of $(V-K)$ versus ($T_{eff}$, [Fe/H]).
The result is $(V-K)_B$= 3.46 $\pm $0.06, corresponding to about $T_{eff}$ = 3950 K.
The associated mass and absolute bolometric magnitude are respectively 0.54 (smaller
than the dynamical mass of about 0.7 to 0.8 \Msun )    
and 8.0.  With a bolometric correction of -1.1 and the transformation from apparent 
to absolute magnitudes this gives V$_B$ = 9.57 too faint by 0.6 magnitude. One possibility
is that 85 Peg B is a spectroscopic binary itself, but we shall leave this hot
subject  for another paper, this one having at least clarified the case of
85 Peg A.

\subsection {$\eta$ Cas} Nothing really new, not already in Fernandes et~al. (\cite{fer98}), 
can be said for this star. The fact that it is more evolved introduces one more parameter 
(age) and prevents to check the unevolved position of the star. The star is 
nevertheless identified in Fig.~\ref{Fig2ab}, and does not raise a particular problem.

\section{How does helium content vary with metallicity?}

Our isochrones are computed with the assumption that the helium content varies
in proportion with metallicity, namely:

\begin{equation} $$Y(Z) = Y_{\rm p} + (Z/Z_{\odot}) \,(Y_{\odot}-Y_{\rm p})$$ \end{equation}

Any unevolved star presenting a significant deviation with respect to the 
isochrone corresponding to its metallicity may indicate that its helium content
deviates from this simple, naive assumption. Actually there is already an 
indication that the Hyades do not follow this rule (Perryman et~al. 
\cite{per98}), Hyades stars being metal-rich by 0.15 dex with respect to the 
Sun, without being correspondingly helium-rich. On the contrary, the metal-rich
stars of the  $\alpha$~Cen system were found also helium-rich (Y=0.30) 
by Fernandes \& Neuforge, (\cite{FN95}) in agreement with  equation (1). 

\begin{figure}[]
\resizebox{\hsize}{!}{\includegraphics{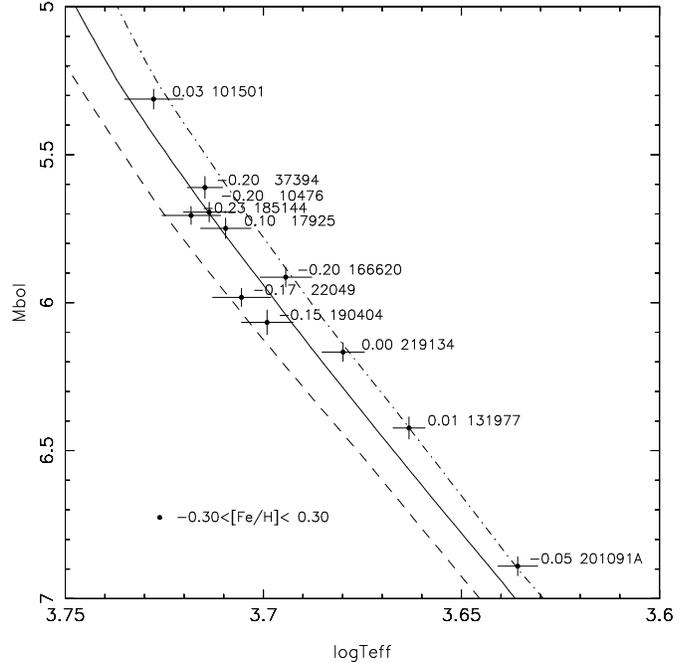}}

\caption{ This figure  is a large scale HR diagram for the subsample of unevolved
stars in the LTE metallicity range -0.3 $<$ [Fe/H] $<$ 0.3. Each star is
labeled by its LTE [Fe/H] value, followed by the HD number of the star.
 The full line is 
a standard isochrone for the  metallicity [Fe/H] = 0. The
dashed line is the standard isochrone corresponding to the LTE  metallicity
of the sample [Fe/H] = -0.3 The dot-dashed line corresponds to an isochrone 
of  metallicity +0.3. All isochrones are for an age of 8 Gyr , but 
evolutionary effects are very small, and age is almost irrelevant.
If stars stand between the limit isochrones,
they do not always closely match the isochrone corresponding to their
metallicity, leaving the impression that $\Delta Y/\Delta Z$ is not constant
in the sample.}
 
\label{Fig6}
\end{figure}

In Fig.~\ref{Fig6} we have plotted the most unevolved stars ($M_{\rm bol} > 5.2$),
labeled by their value of  [Fe/H], with a solar ZAMS and a grid of 
isochrones of 8 Gyr, of metallicities $-0.3$, 0. and $+0.3$. Only stars of
metallicity corresponding to this interval are plotted. We admit that the age of these thin disk
stars is between 0 and 8 Gyr. So if the error bars intersect either the right
metallicity ZAMS or a right metallicity isochrone there is no compelling 
evidence that the scaled helium value is wrong. Globally all stars within 
the metallicity range -0.3 $<$ [Fe/H] $<$ 0.3 are within the corresponding
isochrones. But inside the lane they define, some stars are not located 
exactly where one would expect to see them.
We note that  5 stars
(HD~37394, HD~131977, HD~166620, HD~201091~A, HD~219134) with metallicity very
close to zero or even slightly negative, stand on the  [Fe/H]~=~+0.3, 8 Gyr isochrone. 
The non-LTE correction for their iron abundance is negligible at this metallicity.

As disk stars are younger than halo ones, sedimentation is less important.
Its effect has been estimated, using a 5 Gyr, 0.8 \Msun star of solar composition as a
template. In this star, the shift in the HR diagram due to microscopic
diffusion is of approximately 50K in effective temperature and 0.02 dex in
luminosity. These values are comparable to the observational error bars on
these parameters for individual objects.
Only for the oldest and more massive stars, the  shift can reach more important
values, around 120 K.
Unfortunately, ages are most of the time  unknown for individual {\it unevolved} stars of the disk.
 
Though certainly responsible for a scatter in the HR diagram (due to mass
and age differences), and for a systematic very small shift towards lower
effective temperature, any clear signature of this effect in the  data
seems impossible to decifer presently. Further work is planned on this point. 

Young stars should not show this effect.
Indeed, in another paper (Perrymann et al.1998 , the content in helium of the Hyades,
sufficiently young to forget about diffusion, has been determined, anf found to be 
solar like or slightly below, but certainly not enhanced at the level of the .15
dex level of [Fe/H].   
Also the very young active stars as $\epsilon$~Eridani (HD~22049)
or HD~17925 do not significantly deviate from the solar ZAMS.

 Another possibility for the 5 stars close to the 0 +.3 dex isochrone is 
 that these stars have less helium than the scaled value given by equation (1).
Interpolation of ZAMSs corresponding to various (Y,Z) combinations
could provide an estimate of Y, for a star of known Z. However we must remember that 
the error bar of 0.1 dex on [Fe/H] translates into an error of 0.015 in Y, 
because of the (Y,Z) degeneracy in the HR diagram.
Therefore we have decided to refrain from giving an helium content for individual unevolved stars
until the effects of  sedimentation  are properly quantified at high metallicities.

Summarizing, our data are consistent with the scaled helium content relationship
at a coarse level of accuracy, of the order of a factor of two in metallicity, but 
suggest that a scatter exists at a finer resolution.

\section{Conclusions}

The location in the ($\log T_{\rm eff},\; M_{\rm bol}$) plane of a sample of
over one hundred nearby stars, covering a metallicity range of
 $-1.0~<~{\rm [Fe/H]}~<~+0.3$, with Hipparcos parallaxes of relative accuracy
better than 5 per cent, has been obtained. Most stars have bolometric
magnitudes directly measured, and effective temperatures derived from the
Infra-Red Flux Method, by Alonso et~al. (\cite{alo96a}). The metallicities (LTE)
have been derived from the 1996 edition of the catalogue of ${\rm [Fe/H]}$
determinations by Cayrel de Strobel et~al. (\cite{cayg97}).

The HR diagram shows an unexpected clumping of stars of low metallicity, only
slightly separated from the solar metallicity sequence, as found much earlier,
from ground-based parallaxes by Perrin et~al. (\cite{per77}). However the 
explanation of this fact by a variation of the helium content with metallicity 
$\Delta Y/\Delta Z=5.5$ is no more tenable, with the narrow range of variation 
of Y between $Y_{\rm p}\approx 0.24$ and
$Y_{\odot} \approx 0.28$. A comparison of the data with the standard theoretical
isochrones, computed with improved opacities, show that unevolved stars with
solar metallicities agree more or less with the theoretical expectations,
whereas stars with metallicities below -0.5 dex deviate from the isochrones 
having the same metallicity. The case of $\mu$~Cas, a binary of metallicity 
$-0.7$, is particularly illustrative.

Two effects, recently studied in the literature have been investigated to
explain this observation. The first one concerns only the atmosphere of the
star. Until now, non-LTE abundances for iron were not currently available
in the literature. Th\'evenin \& Idiart (\cite{TI99}) have studied more than
one hundred stars, including $\mu$~Cas and many other metal-poor stars,
and found a non-LTE correction of +0.15 dex for ${\rm [Fe/H]}$ in $\mu$~Cas.
Morel \& Baglin (\cite{MB99}) have studied the microscopic diffusion of helium 
and heavier elements in metal-poor stars of various mass. They 
found that the atmosphere is
depleted in  iron after 10 Gyr, and that the effective temperature of the star
is shifted by about $-30$ to $-200$ K for its luminosity. The combination of
the two effects brings  $\mu$~Cas back on its theoretical sequence. The bulk of
the combined correction is  due to the fact that both the non-LTE 
correction, and the diffusion of heavy elements, make the atmospheric LTE iron 
abundance not representative of the mean inside abundance of iron. 

Among stars with metallicities closer to the solar metallicity there is a global agreement
between the observations and the isochrones with a scaled helium abundance, although
some scatter may be present. New computations of sedimentation at these metallicities
would be necessary to further resolve this issue. 

Also  an enlarged sample of stars, with distances known
with an accuracy of 1 or 2 per cent, appears necessary for a good determination
of the helium content of individual stars. The GAIA mission is very promising
in this respect (Perryman et~al. \cite{per97}). With GAIA the mean precision on 
parallax measurements will be at the 10 micro-arc-second (mas) level up to V=15,
to be compared with 1 mas with Hipparcos up to V=9. In parallel, progresses in 
the analyses of stellar spectra are relevant, as exemplified by the introduction
of departures from LTE of the iron atom.

 More stellar masses are
also deeply needed, not a single mass being known for stars more metal-poor
than $\mu$~Cas.

\begin{acknowledgements}

JF is supported by JNICT (BPD/9919/96). This work was partially supported
by the France-Portugal co-operation (project 059-B0).
We gratefully thank Pierre Morel for his help with CESAM, and   
H.-G. Ludwig for having made available results prior to publication.
MNP is grateful to R. Ca\-na\-vaggia $(\dagger)$ for her help in the  selection
of target stars of the Hipparcos proposal 132 . 
\end{acknowledgements}

\end{document}